\documentclass[prc,showpacs,showkeys,superscriptaddress,nofootinbib,floatfix,onecolumn]{revtex4}
\usepackage[english]{babel}
\usepackage{color}
\usepackage{amsmath}
\usepackage{amsfonts}
\usepackage{amsbsy}
\usepackage{mathrsfs}
\usepackage{graphicx}  
\usepackage{subfigure}
	\newcommand{\eeq}{\end{equation}}
\newcommand{\bea}{\begin{eqnarray}}
	\newcommand{\eea}{\end{eqnarray}}

\def\lsim{\mathrel{\rlap{
			\lower4pt\hbox{\hskip-3pt$\sim$}}
		\raise1pt\hbox{$<$}}}     
\def\gsim{\mathrel{\rlap{
			\lower4pt\hbox{\hskip-3pt$\sim$}}
		\raise1pt\hbox{$>$}}}     

\begin{document}

\title{
Phase space path integral representation of the dynamic structure factor. \\ 
Monte Carlo simulation of strongly correlated  soft-sphere fermions.
} 

\author{V.S.~Filinov}
\author{P.R.~Levashov}
\author{A.S.~Larkin}  
\affiliation{Joint Institute for High Temperatures of the Russian Academy of Sciences, 
	Izhorskaya 13, Bldg 2, Moscow 127412, Russia
}

\begin{abstract}
The dynamic structure factor (DSF) is a mathematical function that contains information about inter-particle correlations 
and their time evolution. Mostly the classical molecular dynamics is used to calculate the DSF of the classical systems. 
On the contrary this article deals with quantum systems and the quantum dynamic structure factor. 
The Wigner formulation of quantum mechanics was used to derive the path integral representation of the DSF, which 
is based on the Wiener-Khinchin theorem showing relation of the the power spectrum of a random paths 
to their correlation function. The $3 {\rm D} $ quantum system of strongly correlated soft-sphere fermions 
was considered as an interesting physical example. 

The developed Wigner path integral Monte Carlo (WPIMC) approach  has been developed to calculate 
the spin--resolved DSFs, the radial distribution functions (RDFs) and other thermodynamic functions  
in a wide range of density and temperatures. 
The physical meaning of the peaks arising on the RDFs  and DSF have been analyzed and explained 
by the manifestation of the interference effects of the exchange and interparticle interactions and, as well  as, 
the wave interference between multiple-scattering. This phenomenon in the system of the soft-sphere scatterers may to be 
the precursor effect of the Anderson localization, which  finds its origin in the wave interference between multiple-scattering paths. 
	\end{abstract} 

\maketitle


\section{Introduction}\label{sec1}
The dynamic structure factor (DSF) is a central quantity describing the physics of quantum many-body systems, capturing structure 
and collective excitations of a system. 
The consideration of the statistical laws like DSF is inevitable as the number of particles in a physical system increases.
Scattering of neutrons,  electrons, and photons  by many-body systems are the most important methods for obtaining information on the spatial
dynamical structures of the many-body system in question.
The spatial structure is derived from elastic scattering, while the excitations of the system can be 
analyzed from the inelastically scattered particles. 
The scattering intensity is essentially determined by the scattering function or dynamic structure factor, which
is the space and time Fourier transform of the density-density correlation function of the unperturbed system.  

Generally speaking a numerical calculation of the DSF of the system requires the solution of the corresponding 
many-body Schrödinger equation to get the knowledge of the many-body eigenstates of the system, which cannot be calculated in general. 
So the another important function that describes the dynamical properties of the system has been introduced. 
It is the density-density correlation function, which is related to the dynamic structure factor.  
Due to the importance of DSF, many works are devoted to the calculation of this quantity. 
There are a lot of successes aided by powerful computational tools, such as  for example, the classical and quantum 
Monte Carlo techniques \cite{ForFilLarEbl,EbelForFil,foulkes2001quantum}, exact diagonalization \cite{noack2005diagonalization}, 
tensor networks \cite{orus2019tensor} and more \cite{baez2020dynamical,mondal2024quantum,roux2013dynamic}. 

An interesting approach involving path integrals was suggested in the article \cite{ferre2016dynamic}, in which
in the path integral formalism quantum particles are presented as ``trajectories'' in the configuration space 
or ``ring polymers'' consisting of a lot of ``beads'' connected by harmonic-like bonds (springs). 
The dynamical structure factor (DSF) in warm dense beryllium  and strongly coupled ions in  plasmas with partially
and strongly degenerate electrons  has been investigated in 
\cite{vorberger2018quantum,li2023ab,moldabekov2019dynamical,tewari1999wavevector}.

The main disadvantage of the path integral Monte Carlo method (PIMC) for simulations of degenerate Fermi systems 
(like electrons) is the ``fermionic sign problem'' 
arising due to the antisymmetrization of a fermion density matrix \cite{feynmanquantum} 
and resulting in thermodynamic quantities being small differences of large numbers associated 
with even and odd permutations. As a consequence, the statistical error in PIMC simulations grows exponentially with the number of particles.  
To overcome this issue many approaches were developed
\cite{zamalin1977monte,EbelForFil,ForFilLarEbl,runeson2018quantum,dornheim2020attenuating}).  
In \cite{ceperley1991fermion,ceperley1992path} to avoid the ``fermionic sign problem'', a restricted fixed--node path--integral 
Monte Carlo (RPIMC) approach was offered. 
In RPIMC only positive permutations are taken into account and the accuracy of the results depends upon the 
conformation of the nodal surface. 

An alternative approach based on the Wigner formulation of quantum mechanics in the phase space 
\cite{wigner1934interaction,Tatr} was used in Refs.~\cite{larkin2017pauli,larkin2017peculiarities} to 
avoid the antisymmetrization of matrix elements and hence the ``sign problem''. This approach allows  
to reproduce the Pauli blocking of fermions and is able to calculate quantum momentum distribution functions
as well as transport properties \cite{EbelForFil,miller2016communication,cotton2016symmetrical,cotton2015symmetrical}. 
Average values of quantum operators in the phase space are also 
available. However, the approach is not applicable at high degeneracy. 
Thus, the ``fermionic sign problem'' for strongly correlated fermions has not been completely solved during the last fifty years.  

In this paper we continue developing the phase space  path integral technique by applying it to the DSF   
of a strongly coupled soft-sphere fermions as example. The developed DSF approximation is antisymmetrized 
and is more accurate and rigorous in comparison with the previous ones \cite{larkin2017pauli,larkin2017peculiarities}. 
In our paper we develop a PIMC method based on the Wigner approach (WPIMC) 
being a compromise between the accuracy and speed of simulations. 

A simple model of 3D soft-sphere fermions is well-known in statistical physics and was chosen to demonstrate the correctness of our method. 
This model includes the one-component plasma (OCP), which is of great astrophysical importance  \cite{luyten1971review,potekhin2010physics}. 
Moreover, the theoretical studies of strongly interacting particles obeying the Fermi--Dirac statistics
is the subject of general interest in many fields of physics, 
in particular, plasma under extreme conditions \cite{EbelForFil}, uniform electron gas  
\cite{dornheim2018uniform}, quantum liquids such as ${\rm ^3He}$ \cite{ceperley1995path} and so on. 
The suggested approach is applicable for predicting DSF not only for bulk structures (3D)
but also for surfaces (2D) in multi-component systems. 

In section~\ref{sec1} we present a path integral description of quantum DSF in the Wigner formulation of quantum mechanics 
and comsider Wiener-Khinchin theorem. 
In section~\ref{htemp} we discuss the path integral representation of the propagator matrix elements. 
In section~\ref{wpint} we consider the path intagral representation of the Wigner functions.
In section~\ref{simulations} we present the results of our simulations by the developed the Wigner path integral 
Monte Carlo method (WPIMC)  for a 3D quantum system of the soft-sphere fermions. 
We consider the spin-resolved RDFs and DSFs of the ideal and  the strongly coupled system of soft-sphere fermions. 
In section~\ref{sec:discussion} we summarize the obtained properties and discuss their physical meaning. 
The derivation of the quantum effective interparticle interaction as well as the details of the WPIMC method 
are given in Appendixes~\ref{sec:appb} and \ref{sec:appc}. 

\section*{Wigner Representation of the Dynamic Structure Factor} \label{sec1}

As an interesting example, we are using the WPIMC for calculations of the  dynamical structure factors (DSF) and  the radial  
distribution function (RDFs) 
for the 3D system of Fermi particles strongly interacting via the soft-sphere potential 
$\phi(r) =\epsilon (\sigma /r)^n$. Here $r$ is the interparticle distance, $\sigma$ characterizes the effective particle size,
$\epsilon$ sets the energy scale  and $n$ is a parameter determining the potential hardness. 
The Hamiltonian of the $N$-particles system ${\hat H}={\hat K}+{\hat U}$ contains the kinetic ${\hat K}$ and the interaction  
${\hat U} = \sum_{i < j}^N \phi(r_{ij})$ energy operators. 
In order to employ quantum simulation of the collective dynamics of density fluctuations over both length 
($k^{-1}$) and time scales ($\omega^{-1}$) we study the dynamic structure factor $S(k, \omega)$  
in dense matter in thermodynamic equilibrium.  
\begin{equation}
	\breve{S}(k, \omega)=Z^{-1} \int {\rm d} t \, 
	\mbox{Tr}\left(e^{-\beta{H}} \hat{B} e^{{\rm i}\hat{H}t}\hat{A}e^{-{\rm i} \hat{H}t} \right)
	e^{-{\rm i} \omega t} \;,
	\label{GFA1}
\end{equation}
where  $\beta =1/k_BT$ is the reciprocal temperature, $\omega$ is in units $( 1/\beta \hbar )$, 
$\rm i$ is the imaginary unit  and $Z\left( N,V,T \right) =\mbox{Tr}\big(e^{-\beta\hat{H}}\big)$ is the canonical 
partition function of the system of $N$ particle in volume $V$. 
For calculation dynamic structure factors $\breve{S}(k, \omega)$ the quantum operators $\hat{B}$ and $\hat{A}$  
have to be chosen in form $\hat{B}=\hat{\rho}(\hat{q},k)= \frac{1}{\sqrt{N}} \sum_{j=1}^N \, e^{-i\langle k|\hat{q}_j\rangle} =\hat{A}^* $. 
In general case operators $B$ and $A$ can be arbitrary quantum operaturs. 

We used it in the symmetric form, 
which may offer certain computational advantages 
\cite{hansen1986theory,van1954correlations,vineyard1958scattering,kubo2012statistical,doll1990equilibrium} 
\begin{eqnarray}
&& 	S(k, \omega)=Z^{-1} \int {\rm d} t \, 
 	\mbox{Tr}\left(\hat{B}e^{{\rm i}\hat{H}t_{\rm c}^{*}}
 	\hat{A}e^{- {\rm i}\hat{H}t_{\rm c}} \right)\, 	e^{-{\rm i} \omega t} = \exp\left(- \frac{\beta \omega}{2}\right)\breve{S}(k, \omega)
 	\nonumber\\&& 
= Z^{-1} \int {\rm d} t \,e^{-{\rm i} \omega t}\, 
\int\hspace{-3pt} {\rm d}\overline{q} \, {\rm d} \overline{\overline{q}} \, {\rm d}\widetilde{q} \,{\rm d} \tilde{\tilde{q}} \, 
\left\langle \overline{q}\left |\hat{B}\right| \overline{\overline{q}} \right\rangle 
\left\langle  \overline{\overline{q}}  \left|e^{{\rm i}\hat{H}t_{\rm c}^{*}}\right|
\widetilde{q} \right\rangle 	\left\langle \widetilde{q} \left |\hat{A}\right| \tilde{\tilde{q}} \right\rangle 
\left\langle  \tilde{\tilde{q}} \left|e^{-{\rm i} \hat{H}t_{\rm c}}\right| \overline{q}  \right\rangle \,   	\;  
 	\label{CFA1}
\end{eqnarray}
where $t$ is time, $t_{\rm c}=t/\hbar-{\rm i} \beta /2$ is  a complex-valued quantity including .
Let us remind that here $\langle q|\tilde{q} \rangle =\delta(q-\tilde{q})$ mean the scalar product of the eigenvectors $|q\rangle $ and 
$|\tilde{q}\rangle $ of the position operator  $\hat{q}$  \cite{Tatr},  
the angular brackets in the expression $\langle q_1|\hat{A}|q\rangle $ mean the scalar product of vectors 
$|q_1\rangle$ and  $|\hat{A}|q\rangle$. 
Henceforth it is convenient to imply  that energy is expressed in units of $k_B T$
($T$ is the temperature of the system), while $q$ and $k$ are a $3N$-dimensional vectors. 

The Wigner representation of the $S(k, \omega)$ can be identically  rewritten in the form, which includes the Weyl symbols of operators  
and generalization of the the Wigner - Liouville function 
\begin{eqnarray}
	S(k, \omega)=(2\pi)^{-12N}\hspace{-4pt} 
	\int\hspace{-3pt} {\rm d}\overline{PQ} \, {\rm d}\widetilde{PQ}
	\,B(\overline{PQ})A(\widetilde{PQ}) \,
	\int {\rm d} t e^{-{\rm i}\omega t}  W\left(\overline{PQ};\widetilde{PQ} ;t \right)  \nonumber \\ 
	=(2\pi)^{-12N}\hspace{-4pt} 
	\int\hspace{-3pt} {\rm d}\overline{PQ} \, {\rm d}\widetilde{PQ}
	\,B(\overline{PQ})A(\widetilde{PQ}) \,
	W\left(\overline{PQ};\widetilde{PQ} ;\omega \right),
	\nonumber\\ \label{CFA}
\end{eqnarray}
where  we have introduced a short-hand notation  
for 6$N$-dimensional phase space points, viz., $\overline{PQ}$ and $\widetilde{PQ}$($\overline{pq}$ and $\widetilde{pq}$), 
with  the momenta and coordinates, respectively, of all the particles in the system. 
Here $B(PQ)$ and $A(PQ)$ denotes the Weyl symbol~\cite{Tatr}  
of the operators $\hat{B}\,$ and  $ \hat{A}$ 
\begin{eqnarray}
&&  B(\overline{PQ})=\int {\rm d} \overline{\xi} \exp(-{\rm i} \langle\overline{P} | \overline{\xi}\rangle)
\left\langle \overline{Q}-\frac{\overline{\xi}}{2}\left |\hat{B}\right|
\overline{Q}+\frac{\overline{\xi}}{2} \right\rangle=
\frac{1}{\sqrt{N}} \sum_{j=1}^N \, e^{-i\langle k|\overline{Q}_j\rangle}=\rho(\overline{Q},k) \;,  
\nonumber\\&& 
A(\widetilde{PQ})=\int {\rm d} \widetilde{\xi} \exp(-{\rm i} \langle \widetilde{P}|\widetilde{\xi}\rangle )
\left\langle \widetilde{Q}-\frac{\widetilde{\xi}}{2}\left |\hat{A}\right|
\widetilde{Q}+\frac{\widetilde{\xi}}{2} \right\rangle=\frac{1}{\sqrt{N}} \sum_{j=1}^N \, e^{i\langle k|\tilde{Q}_j\rangle}
=\rho(\widetilde{Q},k)\;,
\label{WS}
\end{eqnarray} 
Henceforth we introduced the variables $\overline{Q}=(\overline{q}+\overline{\overline{q}} )/2$,  
$\overline{\xi}=(\overline{\overline{q}}-\overline{q})$, 
$\tilde{Q}=(\tilde{q}+\tilde{\tilde{q}} )/2$, $\tilde{\xi}=(\tilde{\tilde{q}}-\tilde{q} )$ and   will use both set of variables.  
Here $W\left(\overline{PQ};\widetilde{PQ} ;t \right)$ is 
\begin{eqnarray}\label{dt4}
&&	W\left(\overline{PQ};\widetilde{PQ} ;t \right) = Z^{-1}
	\int\int {\rm d}\overline{\xi}\, {\rm d}\widetilde{\xi}\, 
	e^{{\rm i} \langle \overline{P}| \overline{\xi}\rangle}e^{{\rm i}\langle\widetilde{P}|\widetilde{\xi}\rangle}
     G\left(\overline{Q\xi};\widetilde{Q\xi} ;t \right) , 
   \nonumber\\&&
G\left(\overline{Q\xi};\widetilde{Q\xi} ;t \right)\equiv 
\left\langle \overline{Q}+\frac{\overline{\xi}}{2} 
\left|e^{{\rm i}\hat{H}t_{\rm c}^{*}}\right|
\widetilde{Q}-\frac{\widetilde{\xi}}{2}  \right\rangle 	
\left\langle \widetilde{Q}+\frac{\widetilde{\xi}}{2} \left|e^{{\rm i} \hat{H}(t^*_{\rm c})}\right|
\overline{Q}-\frac{\overline{\xi}}{2}  \right\rangle^* \, 
\end{eqnarray}

The Wigner function $W\left(\overline{PQ};\widetilde{PQ} ;t \right)$ is presented by the Fourier transforms of the 
``symmetric in time directions''  propagator $G\left(\overline{Q\xi};\widetilde{Q\xi} ;t \right)$. 

Monte Carlo simulations of the DSF can be based on transformation of the DSF according to 
the fundamental Wiener-Khinchin theorem \cite{strichartz2003guide}, 
which relates between the power spectrum density of the stationary random processes and the Fourier transform of their 
autocorrelation function.  So let us introduce the power spectral density (the Fourier transforms) of the both propagator matrix  elements 
in Eq.~(\ref{dt4}) considered further as the complex-valued wide-sense stationary random time processes  
\begin{eqnarray}\label{f1}
	G_1\left(\overline{Q\xi};\widetilde{Q\xi} ;t \right)\equiv 
\left\langle \overline{Q}+\frac{\overline{\xi}}{2} 
\left|e^{{\rm i}\hat{H}t_{\rm c}^{*}}\right| 
\widetilde{Q}-\frac{\widetilde{\xi}}{2}  \right\rangle 
\end{eqnarray}\label{f1}
\begin{eqnarray}\label{f1}
F_1\left(\overline{Q\xi};\widetilde{Q\xi} ;\omega \right)\equiv 
\int {\rm d} t \, \left\langle \overline{Q}+\frac{\overline{\xi}}{2} 
\left|e^{{\rm i}\hat{H}t_{\rm c}^{*}}\right| 
\widetilde{Q}-\frac{\widetilde{\xi}}{2}  \right\rangle  e^{-{\rm i} \omega t}
	&=&2\pi 	\left\langle \overline{Q}+\frac{\overline{\xi}}{2}  
	\left|\delta \left( \omega- \hat{H}\right) e^{-\beta \hat{H} /2} \right| 
	\widetilde{Q}-\frac{\widetilde{\xi}}{2}  \right\rangle \;,
\nonumber
	\end{eqnarray}
and 
	\begin{eqnarray}\label{f2}
G_2\left(\overline{Q\xi};\widetilde{Q\xi} ;t \right)\equiv 
\left\langle \widetilde{Q}+\frac{\widetilde{\xi}}{2} 
\left|e^{{\rm i}\hat{H}t^*_{\rm c}}\right| 
\overline{Q}-\frac{\overline{\xi}}{2}  \right\rangle^*  
\end{eqnarray}\label{f1}
\begin{eqnarray}\label{f1}
F_2\left(\overline{Q\xi};\widetilde{Q\xi} ;\omega \right)\equiv 
\int {\rm d} t \, \left\langle \widetilde{Q}+\frac{\widetilde{\xi}}{2} 
\left|e^{{\rm i}\hat{H}t^*_{\rm c}}\right| 
\overline{Q}-\frac{\overline{\xi}}{2}  \right\rangle^* e^{{\rm i} \omega t}
	&=&  2 \pi  
	\left\langle \widetilde{Q}+\frac{\widetilde{\xi}}{2} 
	\left|\delta \left( \omega - \hat{H}\right) e^{-\beta \hat{H} /2}\right| 
	\overline{Q}-\frac{\overline{\xi}}{2}  \right\rangle 
\nonumber
\end{eqnarray}
Let us note that the dynamic structure factor $S(k, \omega)$ has the form of the Fourier transform of 
the autocorrelation function associated with that processes 
\begin{eqnarray}
&&S(k, \omega)=(2\pi)^{-12N}\hspace{-4pt} 
\int\hspace{-3pt} {\rm d}\overline{PQ} \, {\rm d}\widetilde{PQ}
{\rm d}\overline{\xi}\, {\rm d}\widetilde{\xi}\, 
e^{{\rm i} \langle \overline{P}| \overline{\xi}\rangle}e^{{\rm i}\langle\widetilde{P}|\widetilde{\xi}\rangle}
\rho(\widetilde{Q},k)\rho(\overline{Q},k)
\int {\rm d} t e^{-{\rm i}\omega t}  
G_1\left(\overline{Q\xi};\widetilde{Q\xi} ;t \right)
G_2\left(\overline{Q\xi};\widetilde{Q\xi} ;t \right)
	\nonumber\\&&	
\end{eqnarray}
Making use of the Wiener - Khinchen theorem allows to present the DSF as the power spectral density associated 
with these random processes  
\begin{eqnarray}
S(k, \omega)=(2\pi)^{-12N}\hspace{-4pt} 
\int\hspace{-3pt} {\rm d}\overline{PQ} \, {\rm d}\widetilde{PQ} 
{\rm d}\overline{\xi}\, {\rm d}\widetilde{\xi}\, 
e^{{\rm i} \langle \overline{P}| \overline{\xi}\rangle}e^{{\rm i}\langle\widetilde{P}|\widetilde{\xi}\rangle}
\rho(\widetilde{Q},k)\rho(\overline{Q},k)
\left|F_1\left(\overline{Q\xi};\widetilde{Q\xi} ;\omega \right) 
F_2\left(\overline{Q\xi};\widetilde{Q\xi} ;\omega \right)\right|
\end{eqnarray}

As the Weyl symbols of operator $\hat{\rho}$ does not depend on momentum, then 
we can integrate over $\overline{P}$, $\widetilde{P}$, $\overline{\xi}$, $\widetilde{\xi}$, so  the 
$\overline{\xi}$ and $\tilde{\xi}$ have to be equal to zero due to arising $\delta(\xi)$ in matrix elements of the operators $\hat{B}$  
and $\hat{A}$, so 
\begin{eqnarray}
&&S(k, \omega)=(2\pi)^{-12N}\hspace{-4pt} 
\int\hspace{-3pt} {\rm d}\overline{Q} \, {\rm d}\widetilde{Q}
\rho(\widetilde{Q},k)\rho(\overline{Q},k)
\int {\rm d} t e^{-{\rm i}\omega t}  
G_1\left(\overline{Q0};\widetilde{Q0} ;t \right)
G_2\left(\overline{Q0};\widetilde{Q0} ;t \right)
\nonumber\\&&	
=\int\hspace{-3pt} {\rm d}\overline{Q} \, {\rm d}\widetilde{Q} 
\rho(\widetilde{Q},k)\rho(\overline{Q},k)
\left|F_1\left(\overline{Q0};\widetilde{Q0} ;\omega \right) 
F_2\left(\overline{Q0};\widetilde{Q0} ;\omega \right)\right|  \, 
\end{eqnarray}

In isotropic system after angle averaging DSF has the form 
\begin{eqnarray}
&&	S(|k|,\omega)= 
\frac{8 \pi^3	}{(2\pi)^{6N} N }  
\int\hspace{-3pt} {\rm d}\overline{Q} \, {\rm d}\widetilde{Q}   \, 
\left| \left\langle \overline{Q}  
\left|\delta \left( \omega- \hat{H}\right) e^{-\beta \hat{H} /2} \right| 
\widetilde{Q}  \right\rangle \right| ^2
\sum_{i,j} \frac{\sin (|k||\widetilde{Q}_j-\overline{Q}_i|)}{|k||\widetilde{Q}_j-\overline{Q}_i|}   
\end{eqnarray}

\section{Path integral representation of the propagator matrix elements}\label{htemp}  

The main difficulty of our consideration is that the operators of kinetic and potential enrgy do not commutate and, as a consequence, 
an exact explicit analytical expression for the the propagator matrix elements is unknown.  Nevertheless, it can be constructed using a path integral 
approach~\cite{feynmanquantum,NormanZamalin,zamalin1977monte} 
based on the semi-group property $\exp  \kappa\hat{H} = \exp  \kappa\hat{H}/M \times  \dots \times  \exp  \kappa\hat{H}/M  $ 
($\kappa = t_{\rm c}^{*}/M$, $M$ is a large positive integer), so that  
\begin{multline}
 F_1\left(\overline{Q\xi};\widetilde{Q\xi} ;\omega \right)\equiv 
\int \hspace{-3pt} {\rm d} t 
	\left\langle \overline{Q}+\frac{\overline{\xi}}{2} \left|e^{{\rm i}\hat{H}t_{\rm c}^{*}}\right| 
	\widetilde{Q}-\frac{\widetilde{\xi}}{2}  \right\rangle  e^{-{\rm i} \omega t} 
	=\int \hspace{-3pt} {\rm d} t  e^{-{\rm i} \omega t}  
     \int \hspace{-3pt}  \prod_{j=1}^{M} {\rm d} q_j {\rm d} \underline{q}_j  \\
	\times 	\left\langle \overline{Q}+\frac{\overline{\xi}}{2} \left | e^{-{\rm i} t  (\omega\hat{\rm I }-\hat{H})/M}\right | \underline{q}_1 \right\rangle
    \left\langle \underline{q}_1 \left | e^{ -\beta\hat{H}/2M}\right | \underline{q}_2 \right\rangle 
	\\ \times \left\langle \underline{q}_2 \left | e^{- {\rm i} t (\omega\hat{\rm I }-\hat{H})/M}\right | q_2 \right\rangle  
    \left\langle q_2 \left | e^{-\beta\hat{H}/2M}\right | q_3 \right\rangle
	\\ \times \left\langle q_3 \left | e^{ -{\rm i} t (\omega\hat{\rm I }-\hat{H})/M}\right | \underline{q}_3 \right\rangle  
    \left\langle \underline{q}_3 \left | e^{ -\beta\hat{H}/2M}\right | \underline{q}_4 \right\rangle \dots    
    \\ \times  \left\langle \underline{q}_M \left | e^{- {\rm i} t (-\omega \hat{\rm I }-\hat{H})/M}\right | q_M  \right\rangle 
	\left\langle q_M \left | e^{ -\beta\hat{H}/2M}\right | \widetilde{Q}-\frac{\widetilde{\xi}}{2} \right\rangle . 
	\label{dt2}
\end{multline}

The Weyl symbol of the operator $\hat{H}$ can be presented as the  Hamiltonian function $ H(p,q)$  
\cite{wigner1934interaction,Tatr} 
\begin{eqnarray}
	H(p,q)=\int \hspace{-3pt} {\rm d} \xi \exp ( {\rm i} \left\langle p |  \xi \right\rangle )  
	\left\langle q+\xi/2 \left |\hat{H} \right | q-\xi/2 \right\rangle,
	\label{dt3}
\end{eqnarray}
where the vectors $\xi$ and momentum $p$ are $3N$--dimensional vectors.

The inverse Fourier transform allows to express the matrix elements of operators through their Weyl symbols.
So for large $M$ with the error of the order of $(1/M)^2$ required for the path integral approach \cite{feynmanquantum,zamalin1977monte} we have 

\begin{multline} 
\left\langle \underline{q}_j \left | e^{- {\rm i} t (\omega\hat{\rm I }-\hat{H})/M}\right | q_j  \right\rangle  
\approx \left\langle \underline{q}_j  \left | \hat{\rm I }- \frac{\rm i \omega}{M} \big(E\hat{\rm I }-\hat{H}\big)\right | q_j \right\rangle 
+ {\rm O}\left(\frac{1}{M}\right)^2
\\
=\left(\frac{1}{2\pi}\right)^{(3N)} \int \hspace{-3pt} {\rm d} P_j \exp ( - {\rm i} \left\langle P_j |  \xi_j \right\rangle )
\bigg( 1- \frac{\rm i \omega}{M} \big(E-H(P_j,Q_j)\big) \bigg)  
\\
\approx \left(\frac{1}{2\pi}\right)^{(3N)} \int \hspace{-3pt} {\rm d} P_j \exp\left( - {\rm i} \left\langle P_j |  \xi_j \right\rangle \right)
\exp \left(-\frac{\rm i \omega}{M} \big(E-H(P_j,Q_j)\big) \right)   +  {\rm O}\left(\frac{1}{M}\right)^2,
\label{dt41}
\end{multline}

\begin{figure}[htp] 
	\centering
	\includegraphics[width=0.4\columnwidth,clip=true]{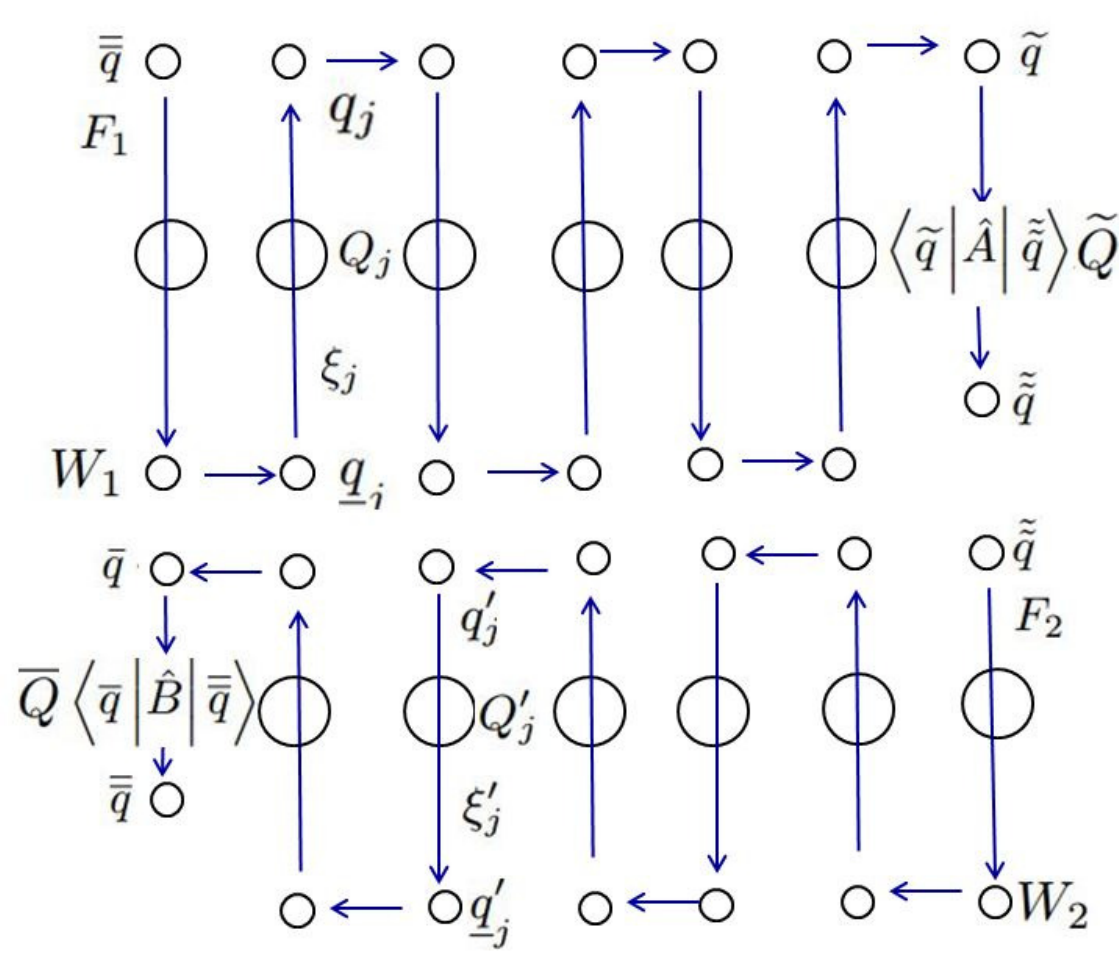}		
	\caption{(Color online)
		The symbolic representation of the DSF and the propagator $G\left(\overline{Q\xi};\widetilde{Q\xi} ;t \right)$ by Eq.~(\ref{dt4}). 
		The ``vertical'' $\left\langle q_j \left | \exp \frac{\rm i t}{M}  \big(\omega \hat{\rm I }-\hat{H}\big)\right | \underline{q}_j \right\rangle$
		and ``horisontal''  $\left\langle q_j \left | \exp -\beta \hat{H}/2M \right | q_{(j+1)} \right\rangle $ 
		($\left\langle \underline{q}_j \left | \exp -\beta \hat{H}/2M  \right | \underline{q}_{(j+1)} \right\rangle $) 
		matrix elements are shown by the related arrows. Matrix elements $ F_1\left(\overline{Q\xi};\widetilde{Q\xi} ;\omega \right) $  
		and $F_2\left(\overline{Q\xi};\widetilde{Q\xi} ;\omega \right)$ correspond to the upper and bottom paths with ``the oppisite 
		time directions''  and with $\left\langle \widetilde{Q}-\frac{\widetilde{\xi}}{2}\left |\hat{A}\right|
		\widetilde{Q}+\frac{\widetilde{\xi}}{2} \right\rangle$ and 
		$\left\langle \overline{Q}-\frac{\overline{\xi}}{2}\left |\hat{B}\right|
		\overline{Q}+\frac{\overline{\xi}}{2} \right\rangle$   respectively   and $M=6$. 
		For the DSF the variables $\bar{\xi}$ and $\tilde{\xi}$ have to be equal to zero due to arising $\delta(\xi)$  
		in matrix elements of the operators $\hat{B}$  and $\hat{A}$ do not depending on momentum. 
		\label{pathh}	}
\end{figure}

The final expression for the product at $\overline{\xi}=0$ and $\widetilde{\xi}=0$  is  equal 
\begin{multline}
	\left\langle \overline{Q} \left | \exp \frac{\rm - i t}{M}  \big(\omega\hat{\rm I }-\hat{H}\big)\right |\underline{q}_1\right\rangle 
	\prod_{j=2}^M \left\langle q_{j} \left | \exp  \frac{-\rm i t}{M} \big(\omega \hat{\rm I }-\hat{H}\big)\right | \underline{q}_j \right\rangle 
		\\
	  \approx \left(\frac{1}{2\pi}\right)^{3NM}  \times {}
	 \prod_{j=1}^{M}  \int \hspace{-3pt}  {\rm d} P_j  \exp ( - {\rm i} \left\langle P_j |  \xi_j \right\rangle )
	\exp \frac{-\rm i t}{M} \big(\omega-H(P_j,Q_j)\big).    
	\label{dt5}  
\end{multline} 
where the new variables 
are defined as $Q_1=(\overline{Q}+\underline{q}_1 )/2$, $Q_j=(\underline{q}_j + q_j)/2$, 
 $\xi_j=(\underline{q}_j - q_j)$ for $j=2, \dots, M$ and 
$H(P_j,Q_j)= \left\langle P_j|P_j \right\rangle/2m + U(Q_j) $ is  the Hamilton function for $N$ paricles at a given $j$. 
Further for convenience we will equally use both set of variables ($Q$, $\xi$) and ($q$, $\tilde{q}$). 
In order to understand Eq.~(\ref{dt2}) its mathematical structure 
is shown in Figure~\ref{pathh} in a symbolic form. 

Then after integration over the time matrix element is presented as 
\begin{eqnarray}
	F_1\left(\overline{Q\xi};\widetilde{Q\xi} ;\omega \right)=
\int \hspace{-3pt} {\rm d} t 
	\left\langle \overline{Q}+\frac{\overline{\xi}}{2} \left|e^{{\rm i}\hat{H}t_{\rm c}^{*}}\right| 
\widetilde{Q}-\frac{\widetilde{\xi}}{2}  \right\rangle  e^{-{\rm i} \omega t} 
=2\pi  \left(\frac{1}{2\pi}\right)^{3NM}  
	\int \hspace{-3pt} {\rm d} P
	\delta (\omega-H(P,Q) \,	W_1\left(P,Q\right)     
	\label{dt6}
\end{eqnarray}	
where $H(P,Q)=\sum_{j=1}^{M} H(P_j,Q_j)/M$, $Q=\{Q_1 , \dots , Q_{M}\}$, $P=\{P_1 , \dots , P_{M}\}$ and $\xi=\{\xi_1, \dots , \xi_{M}\}$ 
are $3NM$--dimensional vectors and  ${\rm d} \underline{q}_1\prod_{j=2}^M{\rm d} q_j {\rm d} \underline{q}_j= {\rm d} Q {\rm d} \xi$, 
and $W_1\left(P,Q\right) $ is the path intagral representation of the Wigner function. 

\section{Path intagral representation of the Wigner functions }\label{wpint} 

The generalization of the  Wigner function $ W_1\left(P,Q\right) $ can be defined  (at $\overline{\xi}=0$ and $\widetilde{\xi}=0$) as 
\begin{multline}
	W_1\left(P,Q\right) = 
\int \hspace{-3pt} {\rm d} \xi e^{ -{\rm i} \left\langle P | \xi \right\rangle } 
\\\times {}
\left\langle \underline{q}_1 \left | e^{-\beta\hat{H}/2M}\right | \underline{q}_2 \right\rangle 
\left\langle q_2 \left | e^{-\beta\hat{H}/2M}\right | q_3 \right\rangle 
\times {}
\left\langle \underline{q}_3 \left | e^{ e^{-\beta\hat{H}/2M}}\right | \underline{q}_4 \right\rangle \cdots
\left\langle q_M \left | e^{ e^{-\beta\hat{H}/2M}}\right | \widetilde{Q}   \right\rangle, 
	\label{dt9}
\end{multline} 
where only ``horizontal'' matrix elements remained (see Fig.~\ref{pathh}). 

In (\ref{GFA1}) we tacitly assumed that the operators the operators $\hat{H}$, $\hat{A}$ and $\hat{B}$ do not depend
on the spin variables. Therefore, summation over spins 
can be safely moved 
here and below, so we do not explicitly mention spin variables, if they are not essential.
However, the spin variables $\sigma$  and the Fermi statistics  can be taken into account by the following 
redefinition of $W_1\left(P,Q\right)$ in the canonical ensemble with temperature $T$  
\begin{widetext}
\begin{multline}
	W_1\left(P,Q\right)  = \frac{1}{Z(\beta)N! } 
	\sum_{\sigma}\sum_{\mathcal{P}} (- 1)^{\kappa_{\mathcal{P}}}  {\cal S}(\sigma,  \mathcal{P} \sigma^\prime)	\big|_{\sigma'=\sigma}\, 
	\int \hspace{-3pt} {\rm d} \xi e^{ -{\rm i} \left\langle P | \xi \right\rangle } 
	\\\times {}
	\left\langle \underline{q}_1 \left | e^{-\beta\hat{H}/2M}\right | \underline{q}_2 \right\rangle 
	\left\langle q_2 \left | e^{-\beta\hat{H}/2M}\right | q_3 \right\rangle 
	\times {}
	\left\langle \underline{q}_3 \left | e^{ e^{-\beta\hat{H}/2M}}\right | \underline{q}_4 \right\rangle \cdots
	\left\langle q_M \left | e^{ e^{-\beta\hat{H}/2M}}\right | \mathcal{P} \widetilde{Q} \right\rangle \\
	= \frac{ 1 }{Z(\beta)N! } \int \hspace{-3pt} {\rm d} \xi \exp ( - {\rm i} \left\langle P | \xi \right\rangle ) \rho^{(1)} \dots \rho^{(M-1)} 
	\sum_{\sigma}\sum_{\mathcal{P}} (-1)^{\kappa_{\mathcal{P}}} 
	{\cal S}(\sigma, \mathcal{P} \sigma^\prime) 	\big|_{\sigma'=\sigma}   \,,
	\mathcal{P} \rho^{(M)}\big|_{ {\mathcal{P}}\widetilde{Q} },
	\label{dt10}
\end{multline}
\end{widetext}  
where the  sum is taken over all permutations $\mathcal{P}$ with the parity $\kappa_{\mathcal{P}}$,    
index $j$ labels the off--diagonal high--temperature density matrices 
$\rho^{(j)}\equiv \langle Q_j \pm  \xi_j/2  |e^{-\frac{1}{M}{\hat H}}| Q_{(j+1)} \pm \xi_{(j+1)}/2 \rangle$. 
Here, as in expression (\ref{dt9}), only the ``horizontal'' matrix elements are present. 

With the error of the order of $1/M^2$ each high--temperature factor can be presented in the form 
$\rho^{(j)}  = \langle Q_j \pm  \xi_j/2  |e^{-\frac{1}{M}{\hat H}}| Q_{(j+1)} \pm  \xi_{(j+1)}/2  \rangle  \approx
e^{-\frac{1}{M}{\hat U(Q_j \pm  \xi_j/2 )}}  \rho^{(j)}_0$  
with $  \rho^{(j)}_0=\langle Q_j \pm  \xi_j /2 |e^{-\frac{1}{M}{\hat K}}| Q_{(j+1)} \pm  \xi_{(j+1)/2} \rangle $, 
arising from neglecting the commutator $\left[K,U\right] / (2M^2)$ and further terms of the expansion.
In the limit $M\rightarrow \infty$ the error of the whole product of high temperature factors 
tends to zero $(\propto 1/M)$
and we have an exact path integral representation of the Wigner functions. 

We imply that momenta and positions are dimensionless variables
$ \tilde{p\lambda}/ \hbar$ and $q/ \tilde{\lambda}$ related to a temperature $MT$  
($\tilde{\lambda}=\sqrt{2\pi\hbar\beta / (m M)}$). 
Spin gives rise to the standard spin part of the density matrix 
${\cal S}(\sigma, \mathcal{P} \sigma^\prime)=\prod_{k=1}^N \delta(\sigma_k,\sigma_{\mathcal{P}k})$, 
($\delta(\sigma_k,\sigma_t)$ is the Kronecker symbol)  with exchange effects accounted for by the permutation
operator  $\mathcal{P}$  acting on coordinates of particles
$\tilde{q}_{(M+1)} $ and spin projections $\sigma'$.

In general the complex-valued integral over $\xi$ in the definition of the Wigner function (\ref{dt10})
can not be calculated analytically. Moreover, this integration is inconvenient for Monte Carlo calculation. 
To overcome this difficulty we have to obtain an explicit expression for $W_1\left(P,Q\right)$. 
However, analytic integration over $\xi$ is possible only for the linear or harmonic potentials, 
when the power of variable $\xi$ is not more than two.  
For this reason we use an approximation for potential energy $U$ arising, for example, from the Taylor expansion up to the first order with respect to $\xi$ 
\cite{LarkinFilinovCPP,larkin2017peculiarities}  
\begin{eqnarray}\label{harmapprox_potential}
	&&U\bigl(( Q_j \pm \xi_j/2  \bigr) 	\approx 
	U\bigl(Q_j\bigr) \pm \frac{1}{2} \left\langle \xi \left|	\frac{\partial U(Q_j)}{\partial Q_j}\right.\right\rangle. 
\end{eqnarray}
Here the second term means the scalar product of the vector $\xi$ and 
the multidimensional gradient of potental energy. 

Then let us replace the variables of integration $Q_j$ by $\zeta_{j} $ 
for any given permutation $\mathcal{P}$ using the substitution
\cite{LarkinFilinovCPP,larkin2017peculiarities}  
\begin{eqnarray}
	Q_j = ( \tilde{P} \tilde{Q} -Q_1)\frac{j-1}{M}+Q_1 + \zeta_{j} ,  
	\label{var}
\end{eqnarray}
where $\tilde{P}$ is the matrix representing the permutation operator $\mathcal{P}$ equal to the unit matrix $E$ with appropriately transposed columns.
This replacement presents each trajectory $Q_j$  as the sum of the ``straight line'' 
$( \tilde{P} \tilde{Q} -Q_1)\frac{j-1}{M}+Q_1 $   and the deviation $\zeta_j $ from it for $j=1, \cdots, M+1$. 
As a consequence the matrix elements can be rewritten  in the form of a path integral 
over \emph{``closed''}  trajectories $ \{  \zeta_{1}, \dots, \zeta_{M} \}$ with $\zeta_{1} =\zeta_{{(M+1)}} =0 $. 

Then after the Hubbard--Stratonovich transformations and some additional ones   
(including analytical continuation of $\phi$ and the integration over $\xi$ and $x$) the main contribution to Wigner function can be written 
in the form containing the  Maxwell distribution with quantum corrections \cite{filinov2022bound,filinov2020uniform,filinov2021monte} 
\begin{widetext}
\begin{multline}
	W_1(P,Q) = 		\frac{C(M)}{Z(\beta)N!}  
	\sum_{\sigma}\sum_{\mathcal{P}} (\pm 1)^{\kappa_{\mathcal{P}}} {\cal S}(\sigma, \mathcal{P} \sigma^\prime)
	\big|_{\sigma'=\sigma}\, 	
	\int { \rm d}  \xi \, 	\exp\Biggl\{ - {\rm i }\langle \xi |P\rangle 	
	-\pi \sum\limits_{j = 1}^M \Biggl[ |Q_{(j+1)}-Q_j|^2 + \frac{1}{4}| \xi_{(j+1)}- \xi_j|^2 \\ 
	{} + (-1)^j \Biggl(\langle Q_{(j+1)}-Q_j |  \xi_{(j+1)} - \xi_j\rangle 
	+ \left\langle  \xi_j \left| \frac{\partial  U^{\tilde{P}}_j }{2M\partial \zeta_j} \right.\right\rangle \Biggr) \Biggr] -U_{\mathcal{P}} \Biggr\} \\
	=\frac{C(M)}{Z(\beta)N!}  
	\sum_{\sigma}\sum_{\mathcal{P}} (\pm 1)^{\kappa_{\mathcal{P}}} {\cal S}(\sigma, \mathcal{P} \sigma^\prime)
	\big|_{\sigma'=\sigma}\, 	
	\times \exp\Biggl\{
	-\pi \frac{|\tilde{P}Q_1-Q_1|^2}{M}  -\sum\limits_{j = 1}^M  \pi | \eta_j |^2  - U_{\mathcal{P}} \Biggr\} 
	\int { \rm d}  x  \exp\Biggl\{ -\sum\limits_{j = 1}^M  \frac{\langle x_j | x_j \rangle}{2}\Biggr\}
	\\ \times  
	\int { \rm d} \xi \exp\Biggl\{ - {\rm i }\langle \xi |P \rangle-i \sum\limits_{j = 1}^M  
	\biggl(  \langle \frac{x_j}{2}+\eta_j|(-1)^j (\xi_{(j+1)} - \xi_j)\rangle - 
	{\rm i }(-1)^j \langle  \xi_j | \frac{\partial  U^{\tilde{P}}_j }{2M\partial \zeta_j} \rangle \biggr) \Biggr\}		
	\\
	\approx \frac{\tilde C(M)}{Z(\beta)N!} 
	\exp\Bigl[ - \sum\limits_{j = 1}^{M}  
	\pi | \eta_j|^2 - U_{\mathcal{E}}  \Bigr]
	\exp\Biggl\{ \frac{M}{4 \pi} \sum\limits_{j = 1}^M  \langle {\rm i } \bar{P}_j | {\rm i } \bar{P}_j  \rangle \Biggr\}
	\times \mathrm{det} \|\tilde{\phi}^{kt} \bigl\|_1^{N/2} \mathrm{det}\bigr\|\tilde{\phi}^{kt} \|_{(N/2+1)}^{N_e} , \,		
	\label{rho-pimc2} 
\end{multline}
\end{widetext}
where $\eta_j \equiv \zeta_{(j+1)} - \zeta_j $, 
\begin{eqnarray} 	
	&&U_{\mathcal{E}} =	\frac{1}{M}	\sum\limits_{j = 1}^M  U_j\biggl(Q_1 + \zeta_j\biggr), \nonumber\\
	&&U_{\mathcal{P}} = \frac{1}{M} \sum\limits_{j = 1}^M   
	U^{\tilde{P}}_j\biggl(( \tilde{P} \tilde{Q} -Q_1)\frac{j-1}{M}+Q_1 + \zeta_{j} \biggr), \nonumber 
\end{eqnarray}
\begin{eqnarray}
	&&\tilde{\phi}^{kt} =  \exp \{-{\pi} \left|r_{kt}\right|^2/M \}  
	\exp \Biggl\{-\frac{1}{M}\sum\limits_{j = 1}^{M} \biggl(
	\bar{\phi}^{kt}_j - \phi^{kt}_j  \biggr) \Biggr\},  \nonumber \\
	&&\bar{\phi}^{kt}_j = \phi \Bigl( \Bigl|r_{tk}\frac{2j}{M} +r_{kt} + (\zeta^k_{j}-\zeta^t_{j})\Bigr|  \Bigr), \nonumber \\
	&&\phi^{kt}_j = \phi \Bigl(\Bigl|r_{kt} + (\zeta^k_{j}-\zeta^t_{j})\Bigr| \Bigr), \nonumber \\
	&&\bar{P_j}\approx P_j - {\rm i}(-1)^j\frac{1}{2M} \frac{\partial  U^{\tilde{P}}_j }{2\partial \zeta_j}, \nonumber
\end{eqnarray}
and $r_{kt}\equiv (\tilde{Q}^k-\tilde{Q}^t)$. The partiail derivatives have here $3N$ components. 
The constants $C(M)$ as well as $\tilde{C}(M)$ are canceled in Monte Carlo calculations. 

In the thermodynamic limit the main contribution in the sum over spin variables comes from the term related
to the equal numbers ($N/2$) of fermions with the same spin projection  \cite{EbelForFil,ForFilLarEbl} and 
the sum over permutations gives  the product of determinants.  
The partition function $Z$ is canceled in Monte Carlo calculations. 

Let us stress that approximation (\ref{rho-pimc2}) has the correct limits in the cases 
of weakly and strongly degenerate fermionic systems. Indeed, in the classical limit the main contribution comes 
from the diagonal matrix elements due to the factor 
$\exp \{-{\pi} \left|r_{kt}\right|^2/M \}$
and potential energies ($U_{\mathcal{E}}$) in the exponents have to be related to indentical permutation. 

At the same time, when the thermal wavelength is of the order of the average interparticle distance 
and the trajectories are highly entangled 
the potential energy weekly depends on permutations and can be approximated by the potential energy ($U_{\mathcal{E}}$) related 
to the identical permutaion 
 \cite{filinov2020uniform, filinov2021monte}. 
  
Similarly we can get the presentation for $F_2\left(\overline{Q\xi};\widetilde{Q\xi} ;\omega \right)$ 
(compare with $F_1$) 
\begin{eqnarray}
F_2\left(\overline{Q\xi};\widetilde{Q\xi} ;\omega \right) = 
	\int \hspace{-3pt} {\rm d} t 
	\left\langle \widetilde{Q}+\frac{\widetilde{\xi}}{2}  \left|e^{{\rm i}\hat{H}t^{*}_{\rm c}}\right| 
	\overline{Q}-\frac{\overline{\xi}}{2}    \right\rangle^*  e^{{\rm i} \omega t} 
		=2\pi  \left(\frac{1}{2\pi}\right)^{3NM}  
	\int \hspace{-3pt} {\rm d} P
	\delta (\omega-H(P^{\prime},Q^{\prime}) \,	W_2\left(P^{\prime},Q^{\prime}\right). 
	\label{dt6}
\end{eqnarray}	
 
Thus, the calculation of the DSF is reduced to WPIMC simulation of the random paths in the phase space 
with probability  $W_1W_2$ (proportional  to the matrix elements of the density matrix ), 
which is similar to the WPIMC simulation of thermodynamic properties. 
Consideration of the DSF  is reduced to calculation  the  of the averaged histogram  of the value 
$\frac{1}{N}\sum_{i,j} \frac{\sin (|k||\widetilde{Q}_j-\overline{Q}_i|)}{|k||\widetilde{Q}_j-\overline{Q}_i|}$ 
versus the internal--energy $\omega$ (in the canonical ensemble) 
\begin{eqnarray}
&&	S(|k|,\omega)= 
\frac{8 \pi^3	}{(2\pi)^{6N} N }  
\int\hspace{-3pt} {\rm d}\overline{Q} \, {\rm d}\widetilde{Q}   \, 
\left|F_1\left(\overline{Q0};\widetilde{Q0} ;\omega \right) 
F_2\left(\overline{Q0};\widetilde{Q0} ;\omega \right)\right| 
\sum_{i,j} \frac{\sin (|k||\widetilde{Q}_j-\overline{Q}_i|)}{|k||\widetilde{Q}_j-\overline{Q}_i|} . 
	\label{GF}
\end{eqnarray}


\section{\label{sec:results}Results of simulations}\label{simulations}  
We present the results of the WPIMC calculations of the radial (RDFs) and the DSF 
for the 3D system of Fermi particles strongly interacting via the soft-sphere potential $\phi(r) =\epsilon (\sigma /r)^n$
 with hardness $n=0.2$. 
Here the density of soft-spheres is characterized by the parameter $r_s=a/\sigma$ 
defined as the ratio of the mean distance between the particles 
$a=\left[3/(4 \pi \tilde{\rho}) \right]^{1/3}$ to $\sigma$  ($\tilde{\rho}$ is the number density). 
The results presented below have been obtained for the following physical parameters
used in \cite{filinov2022solution} for PIMC simulations of helium-3:
$\epsilon \simeq 27$~K, $\sigma \simeq 5.2 \, a_{\rm B}$ ($a_{\rm B}$ is the Bohr radius),
$m_a = 3.016$  is the soft-sphere mass in atomic units.  

To calculate these functions by the WPIMC (see Appendix~\ref{sec:appc} for details) the Markovian chain of particle configurations 
were generated using the Metropolis algorithm. 
We use a standard basic Monte Carlo cell with periodic boundary conditions. 
Between $10^6$ and $3\times 10^6$ equilibrium configurations of 300, 600 and 900 particles represented 
by twenty and forty ``beads'' have been used to calculate average values. 
The convergence and statistical error of the calculated functions were tested with increasing number of Monte Carlo steps,
number of particles and beads at a different hardness of the soft-sphere potential.
It turned out that 600 particles represented by 20 beads were enough to reach good convergence.  

\subsection{Radial distribution functions}\label{rrdf}
Let us start consideration of WPIMC results for RDF and DSF from  ideal system. 
The RDF \cite{kirkwood1935statistical,fisher1964statistical} 
can be written as follows: 
  \begin{eqnarray}  \label{gab-rho}
  	&&g_{ab}(r) =  \int {\rm d} \bar{Q}  \, \delta(|\bar{Q}^a_1-\bar{Q}^b_1|-r)\, 
 \int {\rm d} \tilde{Q} \, {\rm d} P {\rm d}Q \, {\rm d} P^{\prime} {\rm d}Q^{\prime} \,   W_1(P,Q)\,W_2(P^{\prime},Q^{\prime}),
  \end{eqnarray}
where $a$ and $b$ label fermions, with the spin variables, 
The RDF $g_{ab}$ is proportional to the probability density to find a pair of particles of types $a$ and $b$
at a certain distance $r$ from each other. In an isotropic system the RDF depends only on the difference of coordinates because of the
translational invariance of the system.
In a classical non-interacting system of particles $g_{ab} \equiv 1 $, while  
interaction, quantum effects and statistics result in the spacial redistribution of particles and a non-monotonic RDF. 

\begin{figure}[htp] 
	\centering
	\includegraphics[width=0.31\columnwidth,clip=true]{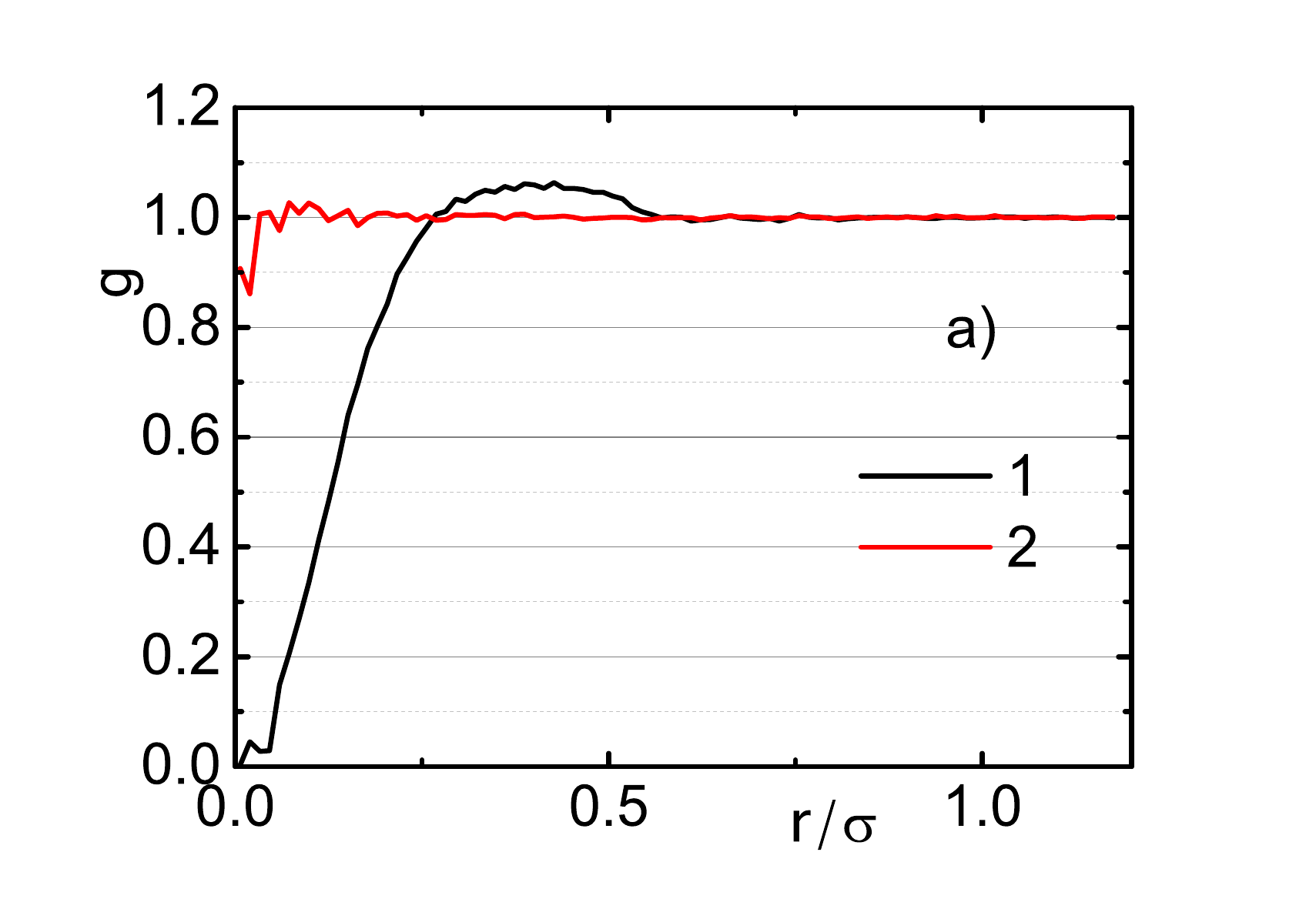}	
	\includegraphics[width=0.31\columnwidth,clip=true]{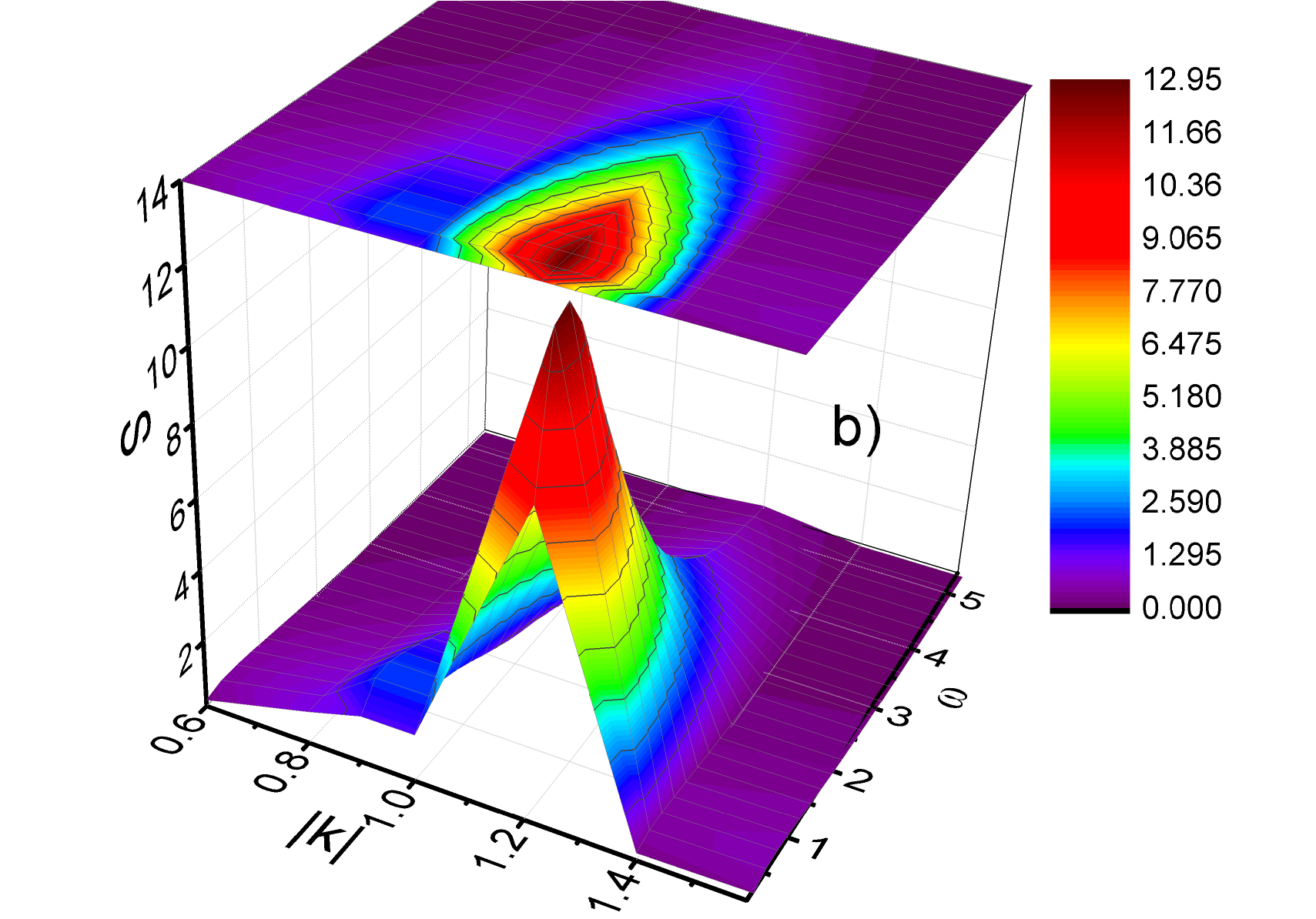}	
	\includegraphics[width=0.31\columnwidth,clip=true]{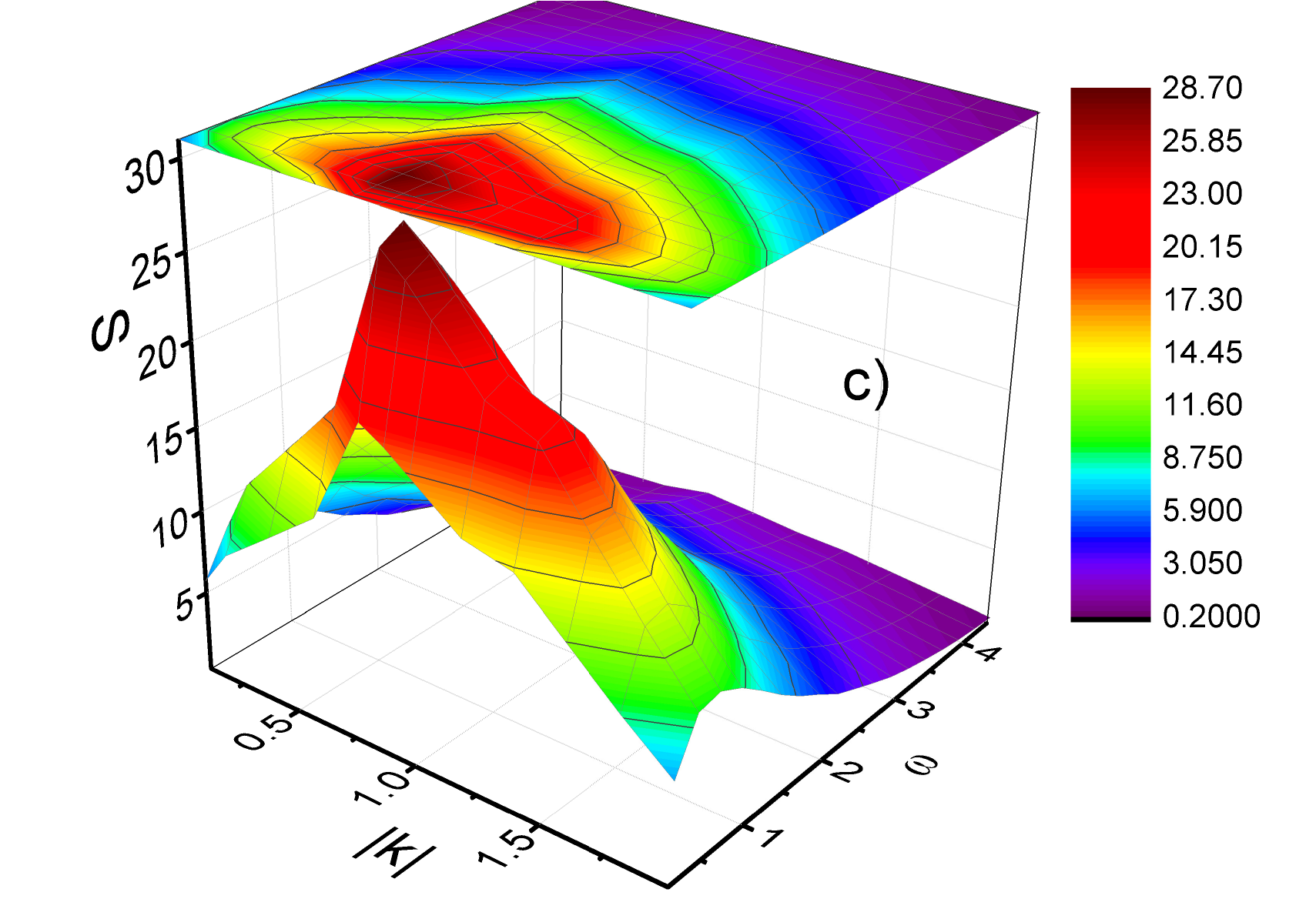}	  	
	\caption{(Color online)The typical spin resolved RDFs for $r_s=2.25$  (panel a) and DSFs for $r_s=2.25$ (panel b) 
		and for $r_s=0.61$  (panel c) of the ideal system of soft-sphere fermions at temperature $T=80$~K and $n=0.2$. 
		Lines for RDF: 1 - the same spin projections, 2 - the opposite spin projections. The DSFs are in conditional units. 
		\label{idmc}
	}
\end{figure} 
The  Figure~\ref{idmc} a) shows RDFs for ideal of soft-sphere fermion system.  
The difference revealed between the RDFs with the same and opposite spin projections of fermions is impressive. 
At small interparticle distances RDFs for opposite spin projection is identically equal to unity contrary to the 
same spin projection ones, which tend to zero due to 
the ``fermi''  repulsion
at distances of the order of the thermal wavelength (lines 2), which is caused by the Fermi statistics effect described 
by the exchange determinant in (\ref{rho-pimc2}). 
This ``fermi''  repulsion leads to the formation of cavities (usually called exchange--correlation holes)
for fermions with the same spin projection 
due to the strong excluded volume effect \cite{barker1972theories} and  arising 
corresponding peak at the distance of the order of the soft sphere thermal wavelength 
($\lambda/\sigma\sim 0.61$, $r_s=2.25$, which is   
less than the average interparticle distance.  
Let us stress that the strong excluded volume effect was observed in the classical systems of hard spheres 
seventy years ago \cite{kirkwood1950radial} and was derived analytically for 1D case in \cite{fisher1964statistical}. 
At large interparticle distances the RDFs tend monotonically to unity. 

\subsection{The dynamic structure factor for ideal system of scatterers }\label{ieddos} 

Analytical estimation of DSF for the point-like uncorrelated scatterers 
with Hamiltonian $\hat{H}=p^2/2m$ ($m$ is the mass of scatterer)  gives the simple exprssion 
\begin{eqnarray}
	S(k,\omega) \propto \exp\left(- \frac{ \omega}{2}\right) \exp(-\frac{1}{4 K } [\omega-K]^2)/(k\lambda)	
	\label{id1}  
\end{eqnarray}
where $K= \frac{(k \lambda)^2 }{4\pi} $, $\lambda$ is the thermal wavelength of scatterer and $K$ and $\omega \hbar  $  are in units $1/\beta $. 
The analytical estimation shows that the maximum intensity occurs 
when energy transfer $\omega$ and momentum transfer k reflect the energy-momentum relation, $\omega=K$ 
of free  nonrelativistic gas particle. Along this line on plane $\omega -  k$ the DSF is decaying proportionally to 
$\exp\left(- \frac{ \omega}{2}\right) $ and inverse power of $k$. 

The WPIMC DSFs for system of the ideal soft-sphere scatterers are presented by the  Figure~\ref{idmc} (panels b) and c)).  
The  Figure~\ref{idmc} b) shows that at  low density and  degeneracy of fermion ($r_s=2.25$, $\lambda/\sigma \approx 0.6$)  
decay of the soft-spheres WPIMC DST maximum  occurs approximately along the line of the fixed $|k|\sigma \approx 1.2$ 
and can be described in the logarithmic scale by straight line, 
what qualitatively agree with analytical estimation for point-like ideal particles. 
Outside this line the DSF oscillate and vanishes exponentially. 
The difference in behavior of the analytical estimation and WPIMC calculations may be attributed to the difference in scattering of the 
pont-like and the soft-sphere scatterers. 

The exponential decay of the WPIMC DSF (Figure~\ref{idmc} (panels b) and c))) versus wavenumber $k$ is accompanied by the 
well pronounced oscillations due to the oscillating nature of the Eq.~(\ref{GF})  . 

The derivative  $\frac{\partial S}{\partial k}$  along the line of 
maximum intensity is equal to zero, while  the derivative  $\frac{\partial S}{\partial \omega}$  may be equal to zero only at points 
of the local maxima. Let us note derivative 
$\frac{\partial\omega}{\partial k}=-\frac{\partial S}{\partial k}/\frac{\partial S}{\partial \omega} $ 
is equal to zero everywhere except may be the points, where we have uncertainty ($\frac{0}{0}$) . This means that the group velocity is 
also equal to zero. The last one is the velocity, with which the overall envelope shape 
of the wave's amplitude (as well as energy and momentum) propagates through space. 
The small group velocity can be attributed to the effect of 
the weak localization arising  in system of randomly distributed in space  scatters.  
This phenomenon in the system of the soft-sphere scatterers is to be the precursor effect of the Anderson localization, which  
finds its origin in the wave interference between multiple-scattering paths. 
The severe interferences can completely halt the waves inside the disordered medium. 

The  more complicated  physical situation takes place at higher density ($r_s=0.61$, see the  Figure~\ref{idmc} c) ). 
Now conditions for the Anderson localization are mostly destroyed and scale of the the DSF  variation  
significantly increased.  The maximum intensity  $S$ on the plane of $k$  and $\omega$ is shifted to $|k|\sim 0.6$, while 
$\omega\sim 0.7$ is practically at the same position as for before for low density.  
Here the average inter particles distance is of order the thermal wave length ($\lambda/\sigma \approx 0.6$) and 
the fermion system is moderately degenerated. 

\subsection{The radial distrubution function for strongly coupled system of scatterers } 
The Figure~\ref{all1} demonstrate the WPIMC RDFs and DSF for a strongly coupled system for 
soft-sphere fermions with hardness $n=0.2$. 
As for ideal system the impressive difference revealed between the RDFs with the same and opposite spin projections of fermions. 
However here at small interparticle distances all RDFs tend to zero not only due to ``exchange'' repulsion of fermions with the same spin projection 
but also due to the  repulsion of the soft--sphere potential.  
An contribution to the repulsion of fermions with the same spin projection at distances of the order of 
the thermal wavelength (lines 2) is caused by the Fermi statistics effect described by the exchange determinant in (\ref{rho-pimc2}).
Now  the exchange determinant in (\ref{rho-pimc2}) depends additionally on the soft-sphere potential  and accounts for the interference effects 
of the exchange and interparticle interactions. 
This combination of repulsions leads to the formation of the high excluded-volume peak on the corresponding RDFs 
and the exchange--correlation cavity for fermions with the same spin projection. 
The RDFs for fermions with the same spin projection show that the characteristic ``size'' of an exchange--correlation cavity  
with corresponding peaks is of the order of the soft sphere thermal wavelength.
($\lambda/\sigma\sim 0.6$, $r_s=2.25$), 
that is less than the average interparticle distance.  
For fermions with the opposite spin projections the interparticle interaction is not enough to form any peakson the RDF. 
(compare lines 1 and 2 in Figure~\ref{all1} a) and b)).  
At large interparticle distances the RDFs tend monotonically to unity due to the short range repulsion of the potential. 
With increasing density the height of the peaks becomes lower. 
Let us remind that these peaks arise due to the interference of the interparticle interaction 
and degeneracy introduced by the determinant in Eq.~(\ref{rho-pimc2}). 

\subsection{The dynamic structure factor for strongly coupled system of scatterers } 

\begin{figure*}[htp] 
	\centering
	\includegraphics[width=0.32\columnwidth,clip=true]{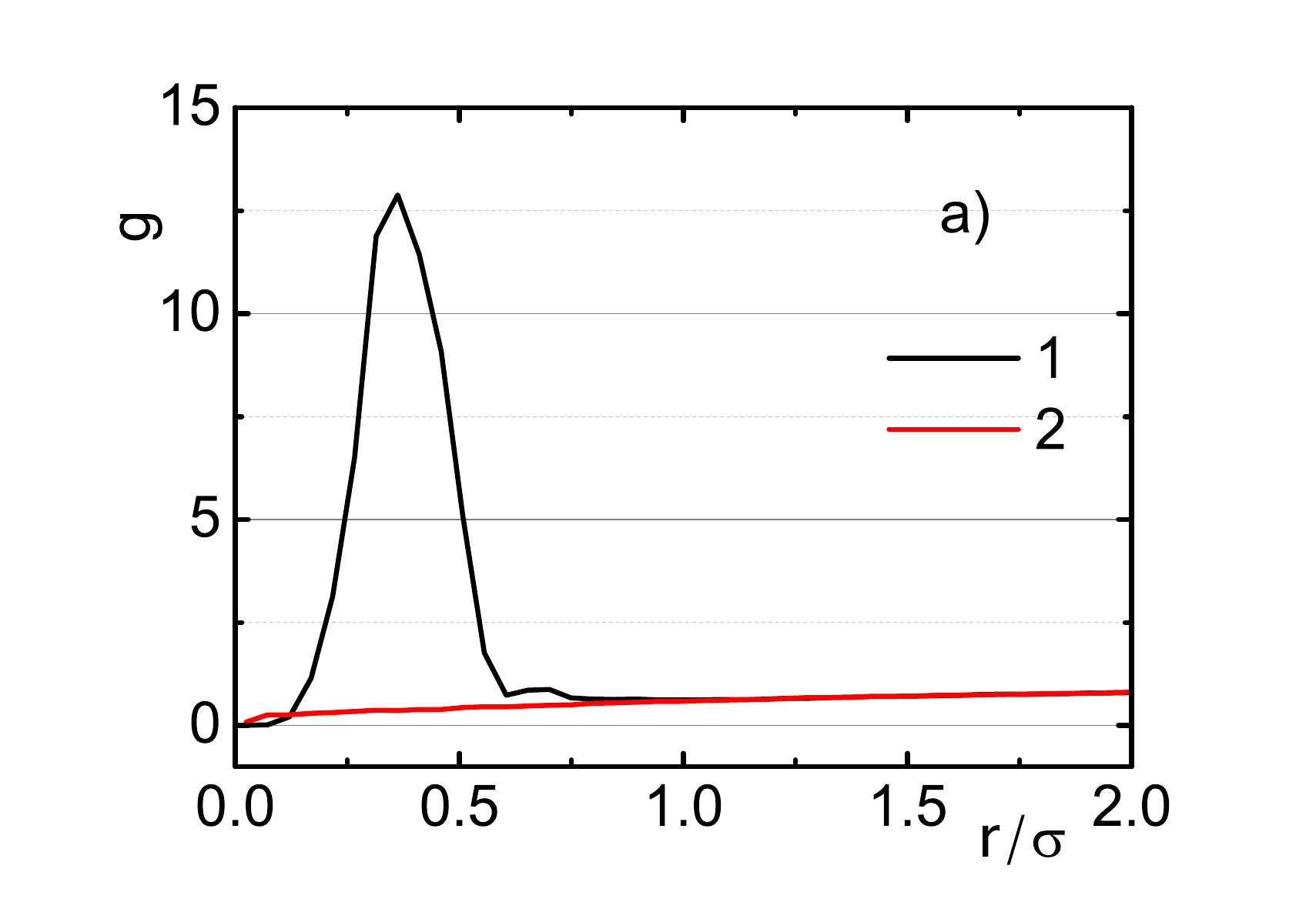}	
	\includegraphics[width=0.32\textwidth,clip=true]{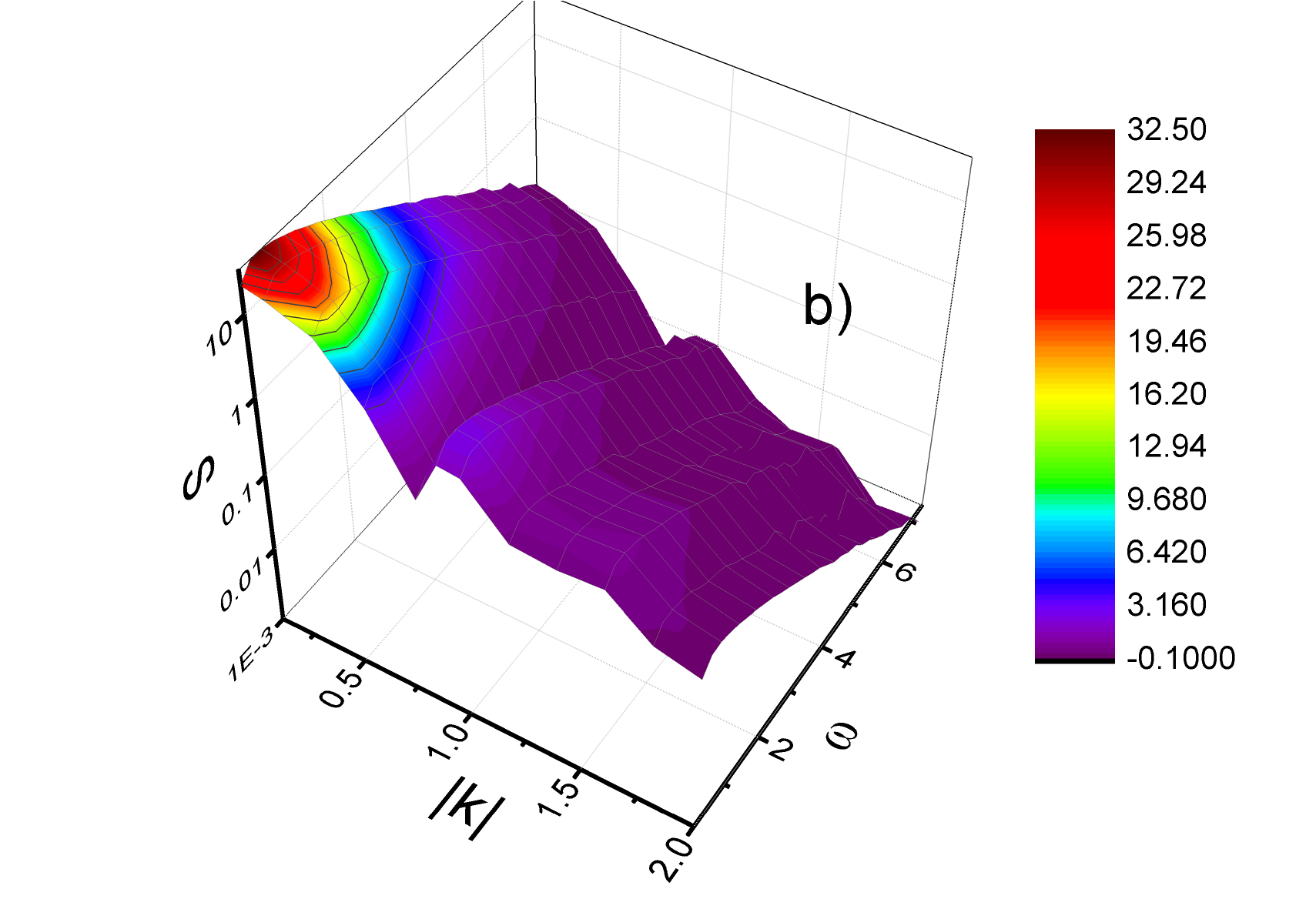}	\\
	\includegraphics[width=0.32\columnwidth,clip=true]{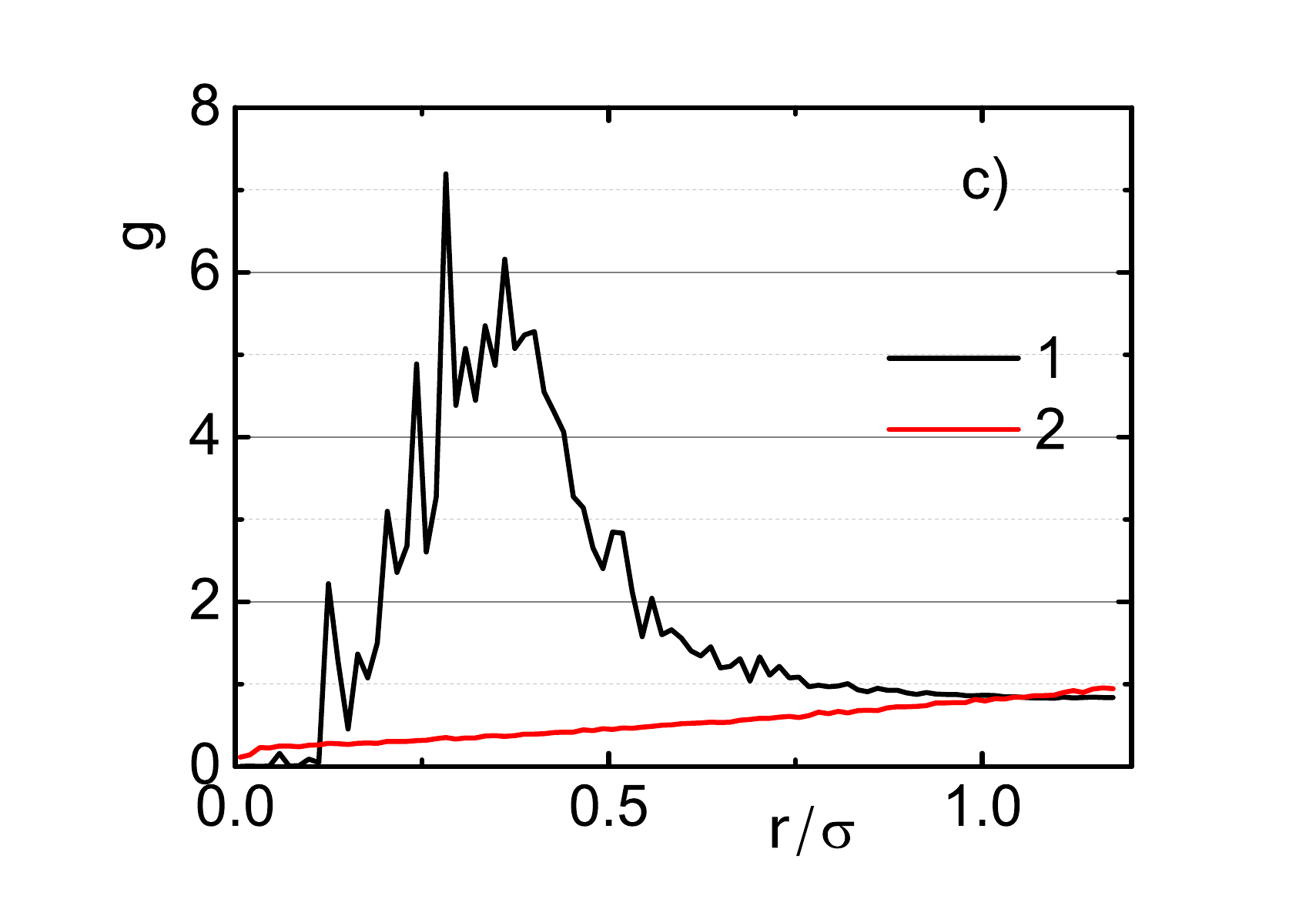}	
	\includegraphics[width=0.32\textwidth,clip=true]{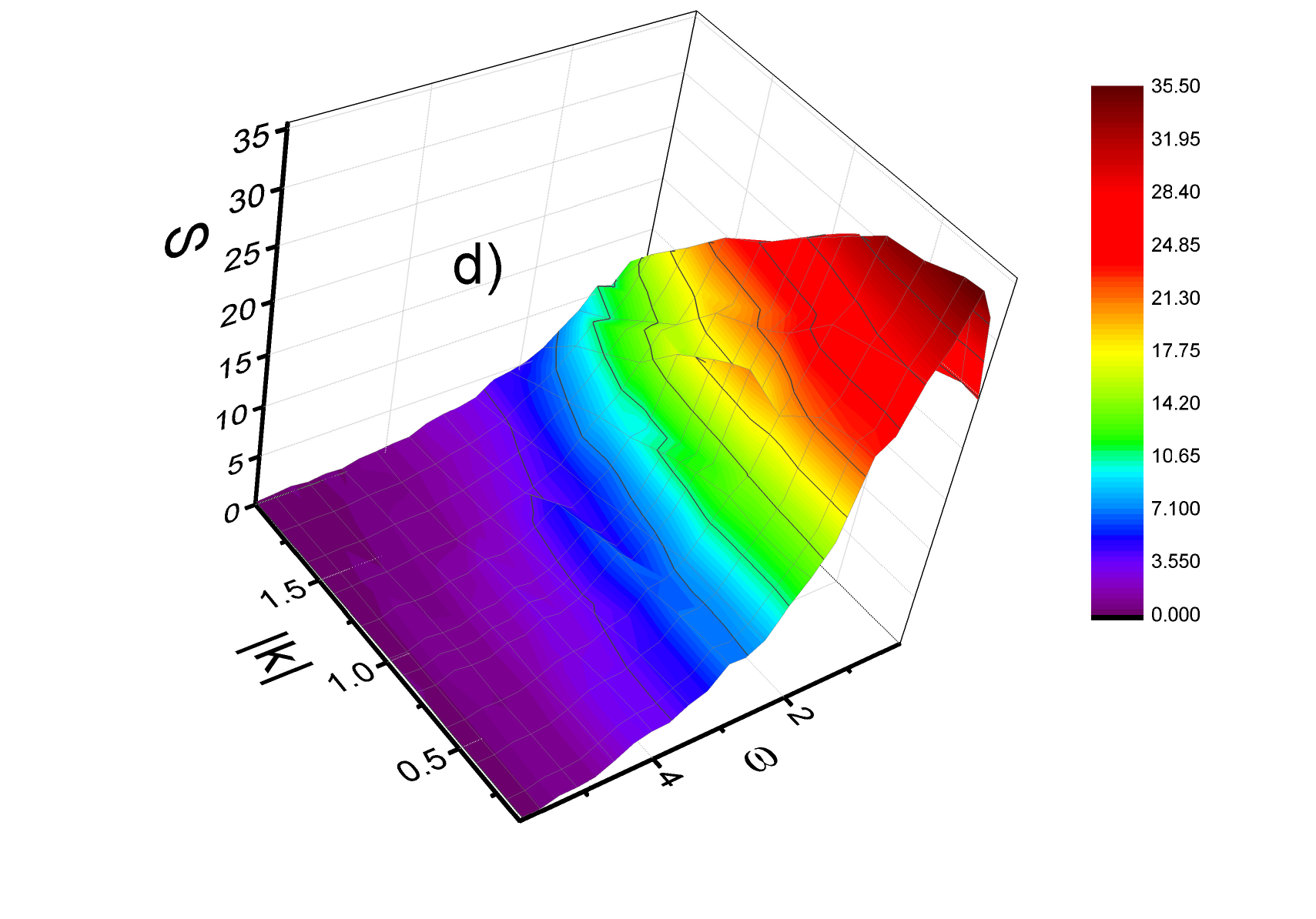}
	\includegraphics[width=0.33\textwidth,clip=true]{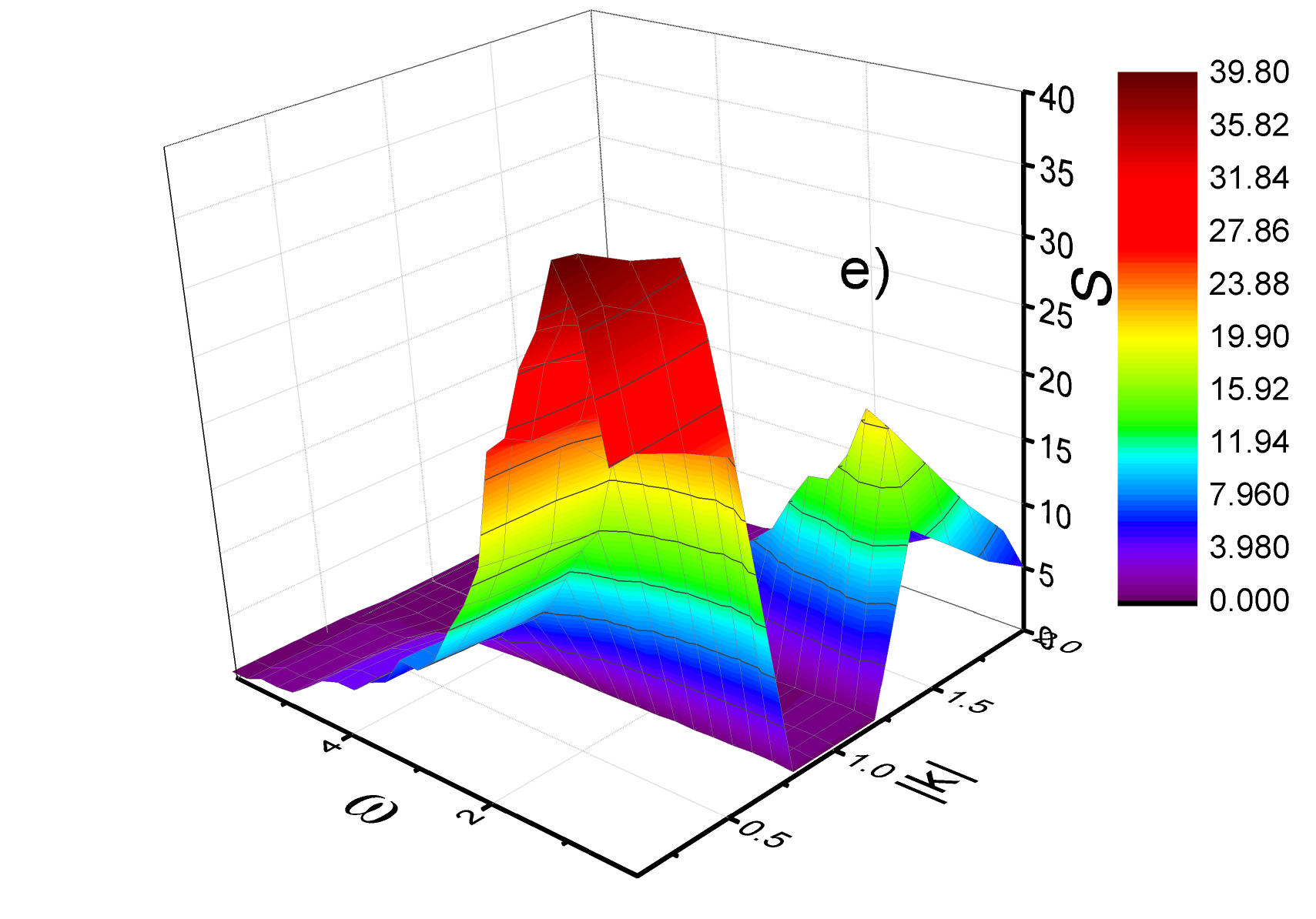}   
	\caption{(Color online)
The WPIMC simulations for $T=20$~K. Density $r_s=2.25$. Panels a) -  RDFs, panel b) - the same spin DSFs. 
Density $r_s=0.61$. Panels c) - RDFs,  panels d) - the same spin DSFs, 
panels e) -  the opposite spin DSF. \\
Lines: 1 - the same spin projections RDFs, 2 - the opposite spin projections RDFs. The DSFs are in conditional units. 
Irregular oscillations indicate the Monte-Carlo statistical error.  \\
		\label{all1}
	}
\end{figure*}

Panels b), d) and e) in Figure~\ref{all1} show pronounced  oscillations of DSF as functios of $k$ for both lower 
($r_s=2.25$) and higher densties ($r_s=0.61$) related to the low and middle fermion degeneracy ($\lambda/\sigma \approx 0.6$ ).  
At lower density ( panel b)) the same and opposite spin DSF practically coincide with each other (not shown) and 
the regions of maximum intensity of the DSF are practically shrinked to the isolated points. The same is valid for the 
opposite spin DSFs (panel d)). 

On the contrary at the same time  the region of maximum intensity of the same spin DSF becomes at higher density ($r_s=0.61$) 
more extended (panel d)).  So the behavior of the DSFs revealed the great difference (see panels e) and d) ). 
Unlike in ideal system the weak dependrnce of the $\omega(k)$ along the line of maximum intensity results (see panel d)) to the  
related smaller derivative  $\frac{d\omega}{dk}$ (group velocity). 
The smaller group velocity can be attributed to the effect of the weak localization arising  in system of the randomly distributed 
in space scatterers. 

The $\omega$ is normalized by temperature equal to $T=20$~K and $T=80$~K  in the  Figures~\ref{idmc} and \ref{all1},  
so the real scales of $\omega$ are different for these two Figures.  

\section{\label{sec:discussion}Discussion} 
The dynamic structure factor is a mathematical function that contains information about inter-particle correlations 
and their time evolution. Experimentally, it can be accessed most directly by inelastic 
neutron scattering or wave scattering.  Mostly the classical molecular dynamics  is used to calculate the DSF and 
so is applied to classical system. However this article deals with quantum systems and the quantum dynamic structure factor. 
The Wigner formulation of quantum mechanics was used to derive the path integral representation of the DSF in the canonical ensemble. 
This representation of the DSF is also based on the Wiener-Khinchin theorem showing how the power spectrum of a random paths in path integrals 
are related to their correlation function. The developed Wigner path integral Monte Carlo (WPIMC) has been developed to calculate 
the spin--resolved DSFs and the radial distribution functions in a wide range of density and temperatures. 

The $3 {\rm D} $ quantum system of strongly correlated soft-sphere fermions for the hardness $n=0,2$ of the interparticle potential 
was considered as an interesting physical example. We present the spin--resolved RDFs ($g$) and DSF of the system of strongly coupled 
soft-sphere fermions for different densities and temperatures.  
The physical meaning of the peaks arising on the RDFs  and DSF have been analyzed and explained 
by the manifestation of the interference effects of the exchange and interparticle interactions and as well  as 
the wave interference between multiple-scattering. This phenomenon in the system of the soft-sphere scatterers may to be 
the precursor effect of Anderson localization, which  finds its origin in the wave interference between multiple-scattering paths. 
The severe interferences can completely halt the waves inside the disordered medium. 

\section*{Acknowledgements}
We thank G.\,S.~Demyanov for value stimulating discussions and the Profs. M. Bonitz, T. Schoof, S. Groth and T. Dornheim for comments. 
The authors acknowledge the JIHT RAS Supercomputer Centre, the Joint Supercomputer Centre of 
the Russian Academy of Sciences, and the Shared Resource Centre ``Far Eastern Computing Resource'' IACP FEB RAS
for providing computing time.

\appendix

\section{\label{sec:appc}Wigner Path Integral Monte Carlo method.}

In this article the  Wigner PIMC method (WPIMC) have been used for calculation in the framework of the approximation (\ref{rho-pimc2}) 
of the Wigner function. In general, to calculate average values of any quantum operators $\langle \hat A\rangle $   
the following representation of $\langle \hat A\rangle $ can be used  \cite{zamalin1977monte,EbelForFil,ForFilLarEbl,larkin2017peculiarities} 
\begin{multline}
	\langle \hat A\rangle  = \int dP dQ  
	A(P,Q) W(P,Q) = {} \\ 
	\frac{\left\langle A(P,Q)
		\cdot h(P,Q\right\rangle _{\tilde{W}}}
	{\langle h(P,Q)\rangle_{\tilde{W}}}, 
	\label{avr1} 	
\end{multline}
where, for example,  the Weyl's symbol of operator $\hat A$ is 
\begin{eqnarray}\label{wigfunc_weylsymbol}
	A(P,Q) = \delta (E- H(P,Q))  \, .
\end{eqnarray}
Here brackets $\bigl\langle g(P,Q)\bigr\rangle _{\tilde{W}}$ denote
the averaging of any function $g(P,Q)$ with a weight $\tilde{W}(P,Q)$ 
\begin{eqnarray}
    \label{mmc_averagesav}
	\langle  g(P,Q) \rangle_{\tilde{W}}  =	\int{dP dQ}\quad g(P,Q)	\tilde{W}(P,Q). 
\end{eqnarray}
To calculate the main contribution to $\langle \hat A\rangle $  the function $\tilde{W}(P,Q)$ 
can be written as the absolute value of a real part of the Wigner functions \cite{Tatr} and a function $h(P,Q)$ 
accounting for the sign of the $\mathrm{Re}(W(P,Q))$ 
\cite{Tatr,EbelForFil,ForFilLarEbl} 
\begin{eqnarray}
  \label{mmc_averageswfh} 
	&&h(P,Q) = \mathrm{sign} (\mathrm{Re}(W(P,Q))), \nonumber \\
	&&\tilde{W}(P,Q) = \left| \mathrm{Re}(W(P,Q)) \right|.
\end{eqnarray}
Note that the partition function $Z$ and constant $\tilde{C}(M)$ in (\ref{rho-pimc2})  are canceled in Monte Carlo calculations. 

The basic idea of a Monte Carlo method is to replace the integration in Eq.~(\ref{mmc_averagesav}) with the averaging over samples $\{\mathbf{\bar{x}}_{1},\mathbf{\bar{x}}_{2},\dots,\mathbf{\bar{x}}_{\tilde{M}}\}$ of a random vector $\mathbf{\bar{x}}$ 
\begin{eqnarray}	
	\langle \hat{A}\rangle=\frac{\sum_{i=1}^{\tilde{M}} A(\overline{\mathbf{x}}_i) h(\overline{\mathbf{x}}_i)  }
	{\sum_{i=1}^N h(\overline{\mathbf{x}}_i)},
	\label{MC3}
\end{eqnarray}  
where the random quantities $\overline{\mathbf{x}}_i\equiv (P,Q)_i$ are drawn from any distribution 
$\tilde{W}(\overline{\mathbf{x}})/\overline{Q} \, $    
($
\overline{Q} = \int_{\Omega}\tilde{W}(\overline{\mathbf{x}})d\overline{\mathbf{x}}   
$).
According to the law of large numbers, if random vectors $\overline{\mathbf{x}}_i$ are not correlated, the statistical error 
is proportional to $1/\sqrt{N}$ and can be estimated using the $3\sigma$-rule. 
If $h(\overline{\mathbf{x}}_i)\equiv 1$ this expression gives the ussual average value. 

The samples $\{\mathbf{\bar{x}}_{1},\mathbf{\bar{x}}_{2},\dots,\mathbf{\bar{x}}_{\tilde{M}}\}$ of a random vector $\mathbf{\bar{x}}$ with a probability 
density $\tilde{W}(\mathbf{\bar{x}})$ can be obtained using the Metropolis algorithm. The Metropolis algorithm is based on the a Markov process,  
which can be constructed by using the transition probabilities. 
This algorithm consists of sequential steps, each of them is divided into two sub-steps: proposal and acceptance. 
Suppose the system is in a state $\mathbf{\bar{x}}_i$, i.e. the random vector $\mathbf{\bar{x}}$ has a value $\mathbf{\bar{x}}_i$.
On the proposal step a new random vector $\mathbf{\bar{x}}'_i$ is generated.
On the acceptance step this new state can be accepted with a probability $A(\mathbf{\bar{x}}_i \to \mathbf{\bar{x}}'_i)$, 
then $\mathbf{\bar{x}}_{i+1} = \mathbf{\bar{x}}'_i$, or rejected, then $\mathbf{\bar{x}}_{i+1} = \mathbf{\bar{x}}_i$.
The acceptance probability $A(\mathbf{\bar{x}}_i \to \mathbf{\bar{x}}'_i)$ must be set to satisfy the detailed balance equation, and the most common choice is
\begin{equation}
	\label{mc_accept}
	A(\mathbf{\bar{x}}_i \to \mathbf{\bar{x}}'_i) = \max\left( 1, \frac{\tilde{W}(\mathbf{\bar{x}}'_i)}{\tilde{W}(\mathbf{\bar{x}}_i)} \right).
\end{equation}
The arising stationary distribution of $\{\mathbf{\bar{x}}_{i}\}$ has to be equal to $\tilde{W}(\overline{\mathbf{x}})$. 


\end{document}